\documentclass{mpe_report}

\usepackage{psfig,graphicx,epsfig}
\usepackage{color}
\usepackage{amsmath,amssymb,epic,eepic,array}

\begin{document}

\pagenumbering{arabic}
\setcounter{page}{60}

\renewcommand{\FirstPageOfPaper }{ 60}\renewcommand{\LastPageOfPaper }{ 63}%\documentclass{mpe_report}

\newcommand{\degrees}[1]{\ensuremath{#1^\circ}}
\newcommand\Eav{{$\left<E\right>$}}

% -----------------------------------------------------------------------------
%\def\R{~ROSAT}
%\def\RAS{\R all sky survey}
% -----------------------------------------------------------------------------
%\begin{document}

\title{The extreme radio emission of PSR B0656+14 ---\\Is B0656+14 a very nearby Rotating Radio Transient?}
\titlerunning{The extreme radio emission of PSR B0656+14}
\author{P. Weltevrede\inst{1} \and  B.~W. Stappers\inst{2,1} \and J.~M. Rankin\inst{3} \and G.~A.~E. Wright\inst{4}}  
\institute{Astronomical Institute ``Anton Pannekoek'', University of Amsterdam, 
Kruislaan 403, 1098 SJ Amsterdam, The Netherlands \and
Stichting ASTRON, Postbus 2, 7990 AA Dwingeloo, The Netherlands \and
Physics Department, 405 Cook Physical Science building, University of Vermont, Burlington, 05405, USA \and
Astronomy Centre, University of Sussex, Falmer, BN1 9QJ, UK}
\maketitle

\begin{abstract}
We present a detailed study of the single radio pulses of PSR
B0656+14. The emission can be characterized by two separate
populations of pulses: bright pulses have a narrow ``spiky''
appearance in contrast to the underlying weaker broad pulses. The
shape of the pulse profile requires an unusually long timescale to
achieve stability (over 25,000 pulses at 327 MHz) caused by spiky
emission. The extreme peak-fluxes of the brightest of these pulses
indicates that PSR B0656+14, were it not so near, could only have been
discovered as an RRAT source. The strongest bursts represent pulses
from the bright end of an extended smooth pulse-energy distribution,
which is unlike giant pulses, giant micropulses or the pulses of
normal pulsars. Longer observations of the RRATs may reveal that they,
like PSR B0656+14, emit weaker emission in addition to the bursts.
\end{abstract}

\section{Introduction}

PSR B0656+14 is one of three nearby pulsars in the middle-age range in
which pulsed high-energy emission has been detected (the ``The Three
Musketeers''). It was included in a recent extensive survey of
subpulse modulation in pulsars in the northern sky at the Westerbork
Synthesis Radio Telescope by Weltevrede et al. 2006a. In the single
pulses analysed for this purpose exceptionally powerful and
longitudinally narrow subpulses reminiscent of ``giant'' pulses were
found.  We therefore set out to explore the full nature of PSR
B0656+14's pulse behaviour in the radio band (Weltevrede et
al. 2006c). This pulsar's extreme bursts are far from typical of older
better-known pulsars, but are similar to those detected in the
recently discovered population of bursting neutron stars. These
Rotating RAdio Transients (RRATs; McLaughlin et al. 2006) typically
emit detectable radio emission for less than one second per day,
causing standard periodicity searches to fail in detecting the
rotation period.  The intermittent extreme bursts we have detected in
PSR B0656+14 led us to argue that this pulsar, were it not so near,
could itself have appeared as an RRAT (Weltevrede et al. 2006b).

\section{Observations}

The results presented in this paper are based on two observations made
using the 305-meter Arecibo telescope. Both observations had a
centre-frequency of 327 MHz and a bandwidth of 25 MHz.  Almost 25,000
and 17,000 pulses with a sampling time of 0.5125 and 0.650 ms were
recorded using the Wideband Arecibo Pulsar Processor (WAPP) for the
observation made in 2003 and 2005 respectively. The Stokes parameters
have been corrected off-line for dispersion, Faraday rotation and
various instrumental polarization effects. For more details we refer
to Weltevrede et al. 2006b and 2006c.

\section{Stability of the pulse profile}

\begin{figure*}[tb]
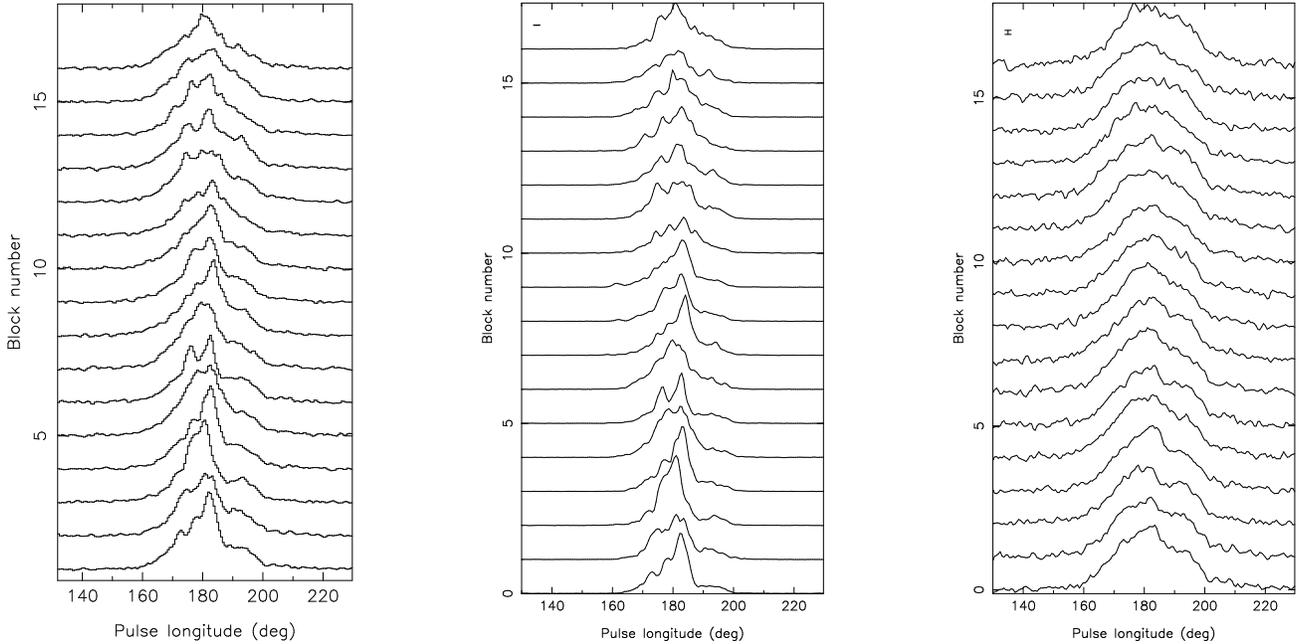

\centerline{\psfig{file=5572f2d.ps,height=8.5cm,clip=} 
  \hspace*{15mm} 
  \psfig{file=5572f13a.ps,height=8.5cm,clip=} 
  \hspace*{15mm} 
  \psfig{file=5572f13b.ps,height=8.5cm,clip=}}
\caption{\label{ProfilesSep}The pulse profiles obtained by averaging
successive blocks of one thousand pulses each of the 2005 Arecibo
observation. The left panel shows all the emission and the middle and
right panel show the spiky and weak emission separately. The 1-sigma
error bars are plotted in the top left corner.}
\end{figure*}

For most pulsars one can obtain a stable pulse profile by averaging a
few to a few hundred pulses, so our observations of up to 25,000
pulses could have been expected to be long enough. However PSR
B0656+14 proved to be far from typical, as the pulse profile is highly
unstable. To illustrate this time dependence, the profiles of
successive blocks of one thousand pulses were calculated (left panel
of Fig. \ref{ProfilesSep}). The scintillation bandwidth is much
smaller than the observing bandwidth, so the intensity of the profiles
are unaffected by interstellar scintillation. This is not because of
systematic errors due to polarization calibration uncertainties
(Weltevrede et al. 2006c).  A much longer observation would be
required to find out if there exists a time scale for the pulse
profile to stabilize.

\section{The spiky emission}

\begin{figure*}[tb]
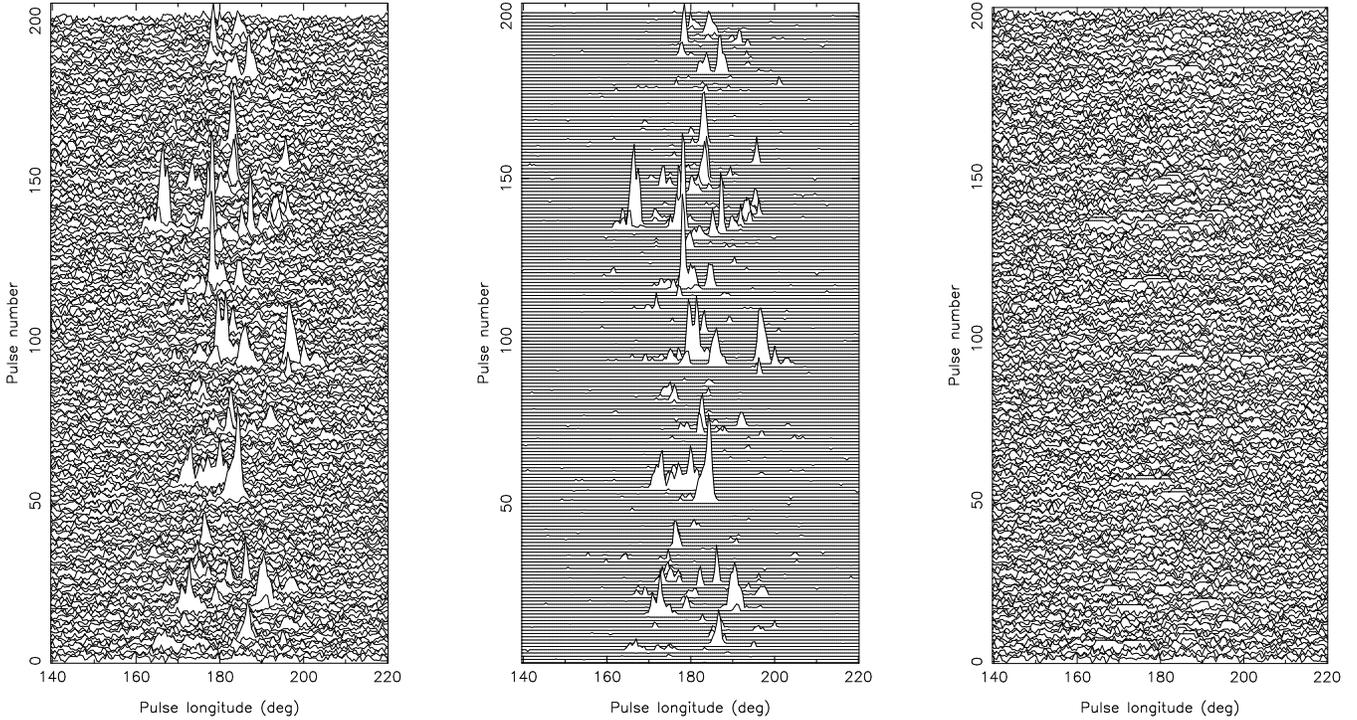

\centerline{\psfig{file=5572f4.ps,height=9.5cm,clip=} 
   \hspace*{8mm} 
   \psfig{file=5572f12a.ps,height=9.5cm,clip=} 
   \hspace*{8mm} 
   \psfig{file=5572f12b.ps,height=9.5cm,clip=}}
\caption{\label{StackSep}A typical sequence of successive pulses (left
panel). The same pulses are shown in the middle and right panel, but
there the emission is separated into the spiky and weak emission
respectively. }
\end{figure*}

A typical pulse sequence of this pulsar is shown in the left panel of
Fig. \ref{StackSep}. One can see that the frequent outbursts of radio
emission are much narrower than the width of the pulse profile. The
emission also has burst-like behaviour in the sense that the radio
outbursts tend to cluster in groups of a few pulse
periods. Furthermore, this clustering sometimes seems to be weakly
modulated with a quasi-periodicity of about 20 pulse periods (see for
instance the bursts around pulse numbers 55, 75, 95, 115 and 135). The
brightest bursts are also shown to have quasi-periodic structures with
a $\sim$11 ms and a $\sim$1-ms timescale (Weltevrede et al. 2006c).
Besides these bursts there are many pulses (and large fractions of the
pulse window) that contain no signal above the noise level. We will
use the term {\em spiky} to refer to these bursts of radio emission.

Although the pulse sequence of the left panel of Fig. \ref{StackSep}
is dominated by the very apparent spiky emission, this is accompanied
by an almost indiscernible background of weak emission.  To separate
these two components of the emission, we have applied an intensity
threshold to the data. The intensities of the time samples in the
pulse stack of the weak emission are truncated if they exceed this
threshold. The time samples in the pulse stack of the spiky emission
contains only samples with intensity in excess of this threshold. When
the pulse stack of the weak emission is added to the pulse stack of
the spiky emission, one retrieves exactly the original pulse stack.

We have set the threshold intensity such that 99\% of the noise
samples are below this threshold value. Not only do the noise
fluctuations make it impossible to completely separate the weak and
spiky emission, it is also very well possible that the energy
distributions of the two components overlap. In Fig. \ref{StackSep}
one can see the pulse stacks obtained after separation of the spiky
and weak emission. The integrated power of this sequence of weak
pulses is about 3 times greater than that of the sequence of the spiky
emission. This shows that a significant fraction of the pulsar's
emission lies at or below the noise level.

The reason why the pulse profile is unstable is the presence of the
spiky emission. These spikes have a very uneven longitude distribution
and therefore many pulses are required to obtain a steady
profile. This is demonstrated in Fig. \ref{ProfilesSep}, where the
profiles of successive blocks of one thousand pulses each are plotted
separately for the spiky and weak emission. One can clearly see that
it is the spiky emission that is highly unstable, whereas the weak
emission converges rapidly to a stable profile.

\section{The radio bursts of PSR B0656+14}

The brightest measured pulse is 116 {\Eav} (where {\Eav} is the
average integrated pulse energy). This is exceptional for regular
radio pulsars and based on the energy of these pulses alone, PSR
B0656+14 would fit into the class of pulsars that emit so-called giant
pulses. Nevertheless, there are important differences between giant
pulses and the bright bursts of PSR B0656+14. The bursts of PSR
B0656+14 have timescales that are much longer than the nano-second
timescale observed for giant pulses, do not show a power-law
energy-distribution, are not confined to a narrow pulse window and are
not associated with an X-ray component. This suggests differing
emission mechanisms for the classical giant pulses and the bursts of
PSR B0656+14. Also the possible correlation between emission of giant
pulses and high magnetic field strengths at the light cylinder clearly
fails for PSR B0656+14. However, giant pulses have been claimed in
other (slow) pulsars that also easily fail this test and for
millisecond pulsars a high magnetic field strengths at the light
cylinder seems to be a poor indicator of the rate of emission of giant
pulses (Knight et al. 2006).

The bursts of PSR B0656+14 are even more extreme when we consider
their peak-fluxes. The highest measured peak-flux of a burst is 420
times the average peak-flux of the pulsed emission, which is an order
of magnitude brighter than the giant micropulses observed for the Vela
pulsar (Johnston et al. 2001) and PSR B1706--44 (Johnston \& Romani
2002). Giant micropulses are not necessarily extreme in the sense of
having a large integrated energy (as required for giant pulses), but
their peak-flux densities are very large. Not only are the bursts of
PSR B0656+14 much brighter (both in peak-flux and integrated energy)
than those found for giant micropulses, they are also not confined in
pulse longitude and they do not show a power-law energy-distribution
as the giant pulses and micropulses do.

\begin{figure*}
\centerline{\psfig{file=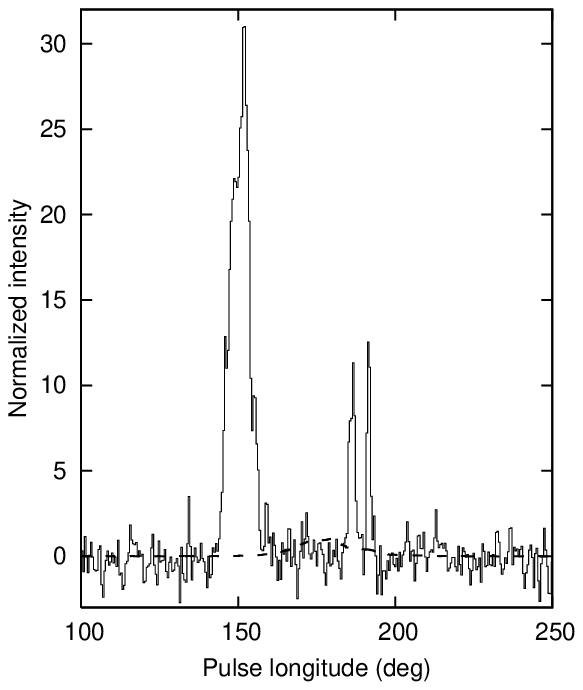,height=5cm,clip=} 
   \psfig{file=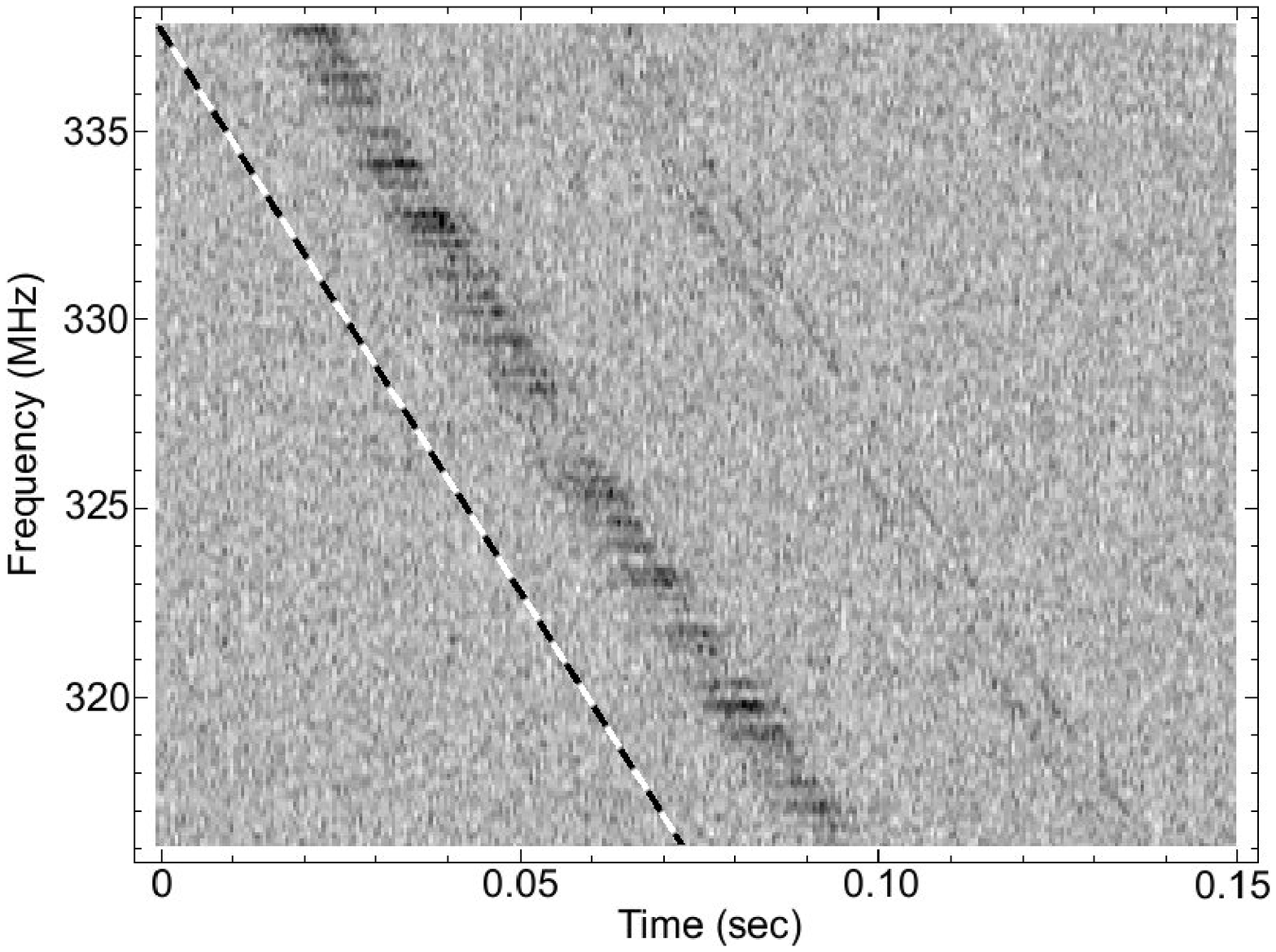,height=4.83cm,clip=} 
   \psfig{file=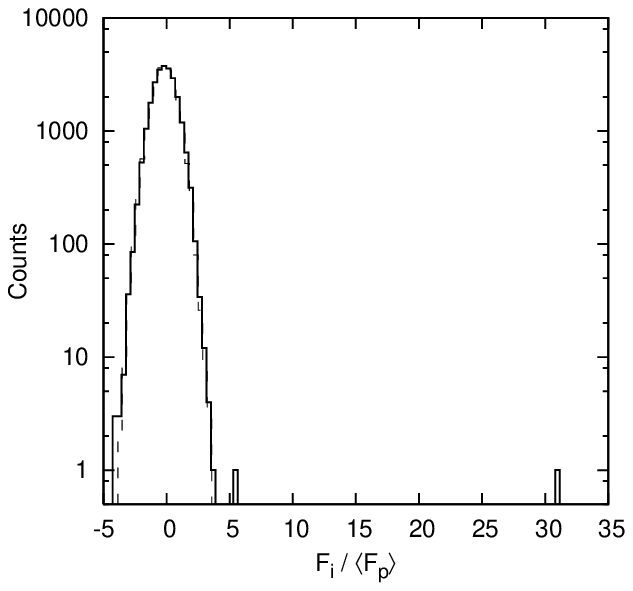,height=5cm,clip=}}
\caption{\label{megapulse}The bright radio burst detected at the
leading edge of the pulse profile in the 2003 observation. {\bf Left:}
The burst (solid line) compared with the average pulse profile (dashed
line). {\bf Middle:} The same burst, but now with frequency
resolution. The data in this plot is not de-dispersed and its
dispersion track matches exactly what is expected for the known
dispersion measure (DM) of this pulsar (dashed line). {\bf Right: }
The longitude-resolved energy-distribution at the longitude of the
peak of the strong pulse (solid line) and the off-pulse distribution
(dashed line). The peak-fluxes ($F_i$) are normalized to the average
peak-flux of the profile ($\left<F_p\right>$).}
\end{figure*}

At the leading edge of the profile we detected a burst with an
integrated pulse-energy of 12.5 {\Eav}. What makes this pulse so
special is that it has a peak-flux that is 2000 times that of the
average emission at that pulse longitude (left panel of
Fig. \ref{megapulse}). Its dispersion track exactly matches what is
expected for this pulsar (middle panel of Fig. \ref{megapulse}),
proving that this radio burst is produced by the pulsar. Notice that
the effect of interstellar scintillation is also clearly visible
(different frequency channels have different intensities) and that the
dispersion track is the same for the two pulses in the centre of the
profile. This burst demonstrates that the emission mechanism operating
in this pulsar is capable of producing intense sporadic bursts of
radio emission even at early phases of the profile. There are only two
bursts with a peak-flux above the noise level detected at the
longitude of the peak of this pulse out of the total of almost 25,000
pulses (see right panel of Fig. \ref{megapulse}). This implies either
that these two bursts belong to an extremely long tail of the
distribution, or that there is no emission at this longitude other
than such sporadic bursts.

\section{The RRAT connection}

The observational facts are that PSR B0656+14 occasionally emits
extremely bright bursts of radio emission which are short in
duration. Although these bursts appears to be different than that of
the giant (micro)pulses, it seems to be similar to those found for the
RRATs. Weltevrede et al. 2006b show that the luminosities of the
bursts of the relatively nearby PSR B0656+14 (288 pc; Brisken et
al. 2003) is very typical for the known RRAT sources.  Although the
slope of the top end of the peak-flux distribution of PSR B0656+14 is
in the range of the giant pulses (between $-2$ and $-3$), it is better
described by a lognormal than by a power-law distribution. This again
suggests that the bright bursts of PSR B0656+14 are different from the
classical giant pulses. The top end of the RRAT distribution with the
highest number of detections seems to be harder (with a slope $-1$),
but for the other RRATs this is as yet unclear. For instance, the tail
of the distribution of PSR B0656+14 seems to be consistent with the
distribution of the RRAT with the second highest number of detections.

Normal periodicity searches failed to detect the RRATs, which places
an upper limit on the average peak-flux density of weak pulses among
the detected bursts of about 1:200 (McLaughlin et al. 2006). Because
the brightest burst of PSR B0656+14 exceeds the underlying peak-flux
by a much greater factor, PSR B0656+14 could have been identified as
an RRAT, were it not so nearby. Were it located twelve times farther
away (thus farther than five of the RRATs), we estimate that only one
burst per hour would be detectable (the RRATs have burst rates ranging
from one burst every 4 minutes to one every 3 hours). The typical
burst duration (about 5 ms) of PSR B0656+14 also matches that of the
RRATs (between 2 and 30 ms).

Were PSR B0656+14 twelve times more distant, the RMS of the noise
would increase by a factor 144 relative to the strength of the pulsar
signal. When we artificially add gaussian-distributed noise at this
level to the observation, we find no sign of the pulsar's (2.6-Hz)
rotation frequency in 35-minute segments of the data yet the brightest
pulse is easily detected with $18\sigma$ (Weltevrede et
al. 2006b). For telescopes with a lower sensitivity than Arecibo
(e.g. Parkes) then even if PSR B0656+14 were quite a bit closer it
would not reveal it's periodicity in a similarly long observation.
Only in the spectrum of the whole 1.8-hour observation the periodicity
of a twelve times more distant PSR B0656+14 would be marginally
detectable for Arecibo.  This means that a distant PSR B0656+14 could
only be found as a RRAT in a survey using Arecibo, unless the
pointings were unusually long.

\section{Implications and discussion}

The emission of PSR B0656+14 can be characterized by spiky (with low
occurrence rate within each pulse) and weak emission (with a high
occurrence rate over the full width of the pulse).  PSR B0656+14
intermittently emits pulses that are extremely bright compared to
normal pulsars and with pulse energies well above ten times the
average pulse-energy these pulses formally qualify as giant
pulses. Nevertheless these pulses differ from giant pulses and giant
micropulses in important ways.  Many of the exceptional properties of
PSR B0656+14 have led us to point out that this pulsar, were it not so
near, could have been discovered as an RRAT.

Our identification of PSR B0656+14 with RRATs implies that at least
some RRATs could be sources which emit pulses continuously, but over
an extremely wide range of energies. This is in contrast to a picture
of infrequent powerful pulses with otherwise no emission. Therefore,
if it indeed turns out that PSR B0656+14 (despite its relatively short
period) is a true prototype for an RRAT, we can expect future studies
to demonstrate that RRATs emit much weaker pulses among their
occasional bright bursts. We would also predict that their integrated
profiles will be found to be far broader than the widths of the
individual bursts, and will need many thousands of bursts to
stabilize.

% Example list of References

%\end{document}

  \clearpage

\end{document}